\documentclass[12pt]{article}%
\usepackage{amsmath,latexsym}
\usepackage{graphicx}
\usepackage{amsmath}
\usepackage{amsfonts}
\usepackage{amssymb}%
\begin{document}

\title{{The no-boundary proposal via the five-dimensional
   Friedmann-Lema\^{i}tre-Robertson-Walker model}}
   \author{Peter K.F. Kuhfittig* and Vance D. Gladney*\\
  \footnote{kuhfitti@msoe.edu}
 \small Department of Mathematics, Milwaukee School of
Engineering,\\
  \small Milwaukee, Wisconsin 53202-3109, USA}

\date{}
 \maketitle
\begin{abstract}\noindent
Hawking's proposal that the Universe has no temporal
boundary and hence no beginning depends on the notion
of imaginary time and is usually referred to as the
\emph{no-boundary proposal}.  This paper discusses
a simple alternative approach by means of the
five-dimensional
Friedmann-Lema\^{i}tre-Robertson-Walker model.
\\
\\
PAC numbers: 04.20.-q, 04.50.+h
\\
\end{abstract}

 \noindent
\section{Introduction}
It is well known that Stephen Hawking introduced the
no-boundary proposal using the concept of imaginary
time, making up what is called Euclidean spacetime
\cite{HH83}.  While ordinary time would still have
a big-bang singularity, imaginary time avoids this
singularity, implying that the Universe has no
temporal boundary and hence no beginning.  By
eliminating the singularity, the Universe
becomes self-contained in the sense that one
would not have to appeal to something outside
the Universe to determine how the Universe
began.

Given that imaginary time is orthogonal to
ordinary time, one could view imaginary time as
an extra dimension.  The existence of an extra
dimension suggests a different starting point,
the five-dimensional
Friedmann-Lema\^{i}tre-Robertson-Walker model
(FLRW) \cite{TS}.  This model produces a simple
alternative version of the no-boundary proposal.

\section{The FLRW model}

Let us recall the FLRW model in the usual four
dimensions \cite{Wald}:
\begin{equation}\label{E:line1}
  ds^2=-dt^2+a^2(t)\left [\frac{dr^2}{1-Kr^2}+r^2
  (d\theta^2+\text{sin}^2\theta\,d\phi^2)\right ],
\end{equation}
where $a^2(t)$ is a scale factor.  Here $K=1/R^2$,
where
\begin{equation}\label{E:R}
   R=\sqrt{x^2+y^2+z^2+w^2};
\end{equation}
also, $K>0$, $K=0$, and $K<0$ (imaginary $R$)
correspond to a closed, flat, and open universe,
respectively.  For $K>0$, the substitution
$r=\frac{1}{\sqrt{K}}\text{sin}\,\psi$ yields
\begin{equation}\label{E:line3}
   ds^2=-dt^2+a^2(t)R^2[d\psi^2+\text{sin}^2\psi(d\theta^2
  +\text{sin}^2\theta\,d\phi^2)].
\end{equation}
The spatial part of the metric is a 3-sphere, having
neither a center nor an edge.  Here we need to recall
that a 3-sphere can be defined as a three-dimensional
boundary of a fictitious four-dimensional ball of
radius $R$, as defined in Eq. (\ref{E:R}).  In the
discussion below we will assume a unit sphere,
i.e., $R=1$.

Returning to line element (\ref{E:line3}), observe
that the singularity in line element (\ref{E:line1})
has been removed, showing that we are dealing with
a coordinate singularity, not a physical one.  The
case $K<0$ does not lead to a singularity.  Here
the substitution $r=\frac{1}{\sqrt{|K|}}\text{sinh}
\,\psi$ leads to the analogous line element in
hyperbolic coordinates (with $R=1$):
\begin{equation}\label{E:line4}
   ds^2=-dt^2+a^2(t)[d\psi^2+\text{sinh}^2\psi(d\theta^2
  +\text{sin}^2\theta\,d\phi^2)].
\end{equation}

Turning now to the five-dimensional spatially
homogeneous and isotropic FLRW metric, we have,
according to  Ref. \cite{TS},
\begin{equation}\label{E:line5}
   ds^2=-dt^2+a^2(t)\left \{\frac{dr^2}{1-Kr^2}
   +r^2[d\psi^2+\text{sin}^2\psi(d\theta^2
  +\text{sin}^2\theta\,d\phi^2)]\right \}.
\end{equation}
The above substitutions for $r$ now lead to the
respective line elements
\begin{equation}\label{E:line6}
   ds^2=-dt^2+a^2(t)\left \{d\chi^2
   +\text{sin}^2\chi[d\psi^2+\text{sin}^2\psi(d\theta^2
  +\text{sin}^2\theta\,d\phi^2)]\right \}.
\end{equation}
and
\begin{equation}\label{E:line7}
   ds^2=-dt^2+a^2(t)\left \{d\chi^2
   +\text{sinh}^2\chi[d\psi^2+\text{sin}^2\psi(d\theta^2
  +\text{sin}^2\theta\,d\phi^2)]\right \}.
\end{equation}
The respective spatial parts can be viewed as
four-dimensional boundaries of a five-dimensional
unit sphere and a five-dimensional unit hyperboloid.

Ref. \cite {TS} makes the usual assumption that any
extra spatial dimension has been compacted to small
size in the course of the evolution of the Universe.
Since the extra dimension does not participate in
the expansion, more realistic line elements are
obtained by leaving $a^2(t)$ in the original
position:
\begin{equation}
   ds^2=-dt^2+d\chi^2
   +\text{sin}^2\chi\left\{a^2(t)[d\psi^2+\text{sin}^2\psi(d\theta^2
  +\text{sin}^2\theta\,d\phi^2)]\right \}
\end{equation}
or
\begin{equation}
   ds^2=-dt^2+d\chi^2
   +\text{sinh}^2\chi\left\{a^2(t)[d\psi^2+\text{sin}^2\psi(d\theta^2
  +\text{sin}^2\theta\,d\phi^2)]\right \}.
\end{equation}

\section{{\textbf{The no-boundary proposal}} }

Stephen Hawking's proposal that the Universe had no
beginning \cite{HH83} depends on the notion of
imaginary time.  If one thinks of ordinary time as a
real axis pointing to the future in one direction and
the past in the other, then the imaginary-time axis is
perpendicular to the real-time axis.   The main idea
in this note is to show that the existence of an
extra spatial dimension can also lead to the
no-boundary proposal.

Since our approach is heavily dependent on the
time-component, we first recall that in geometrized
units ($G=c=1$) the Schwarzschild line element
\begin{equation*}
  ds^2=-\left(1-\frac{2M}{r}\right)dt^2+\frac{dr^2}{1-2M/r}
  +r^2(d\theta^2+\text{sin}^2\theta\,d\phi^2)
\end{equation*}
describes a black hole having an event horizon at $r=2M$.
Ordinarily, $r>2M$.  Inside the event horizon we have
$r<2M$, so that the first two terms undergo sign changes.
So $t$ becomes spacelike and $r$ timelike, leading to
the traditional argument that motion in the $r$-direction
cannot be reversed, so that escape from a black hole is
impossible.

To show that the time component can also become spacelike
in the present study, let us recall the Lorentzian metric
\[
  ds^2=-c^2dt^2+dx^2+dy^2+dz^2\quad (\text{signature:} -+++)
\]
or
\[
  ds^2=c^2dt^2-dx^2-dy^2-dz^2\quad (\text{signature:} +---)
\]
temporarily reintroducing $c$.  (The choice of signature is
merely a matter of convenience.)  In spherical coordinates,
\[
    ds^2=-c^2dt^2+dr^2+r^2(d\theta^2
    +\text{sin}^2\theta\,d\phi^2)
\]
or
\[
    ds^2=c^2dt^2-dr^2-r^2(d\theta^2
    +\text{sin}^2\theta\,d\phi^2).
\]
In the Minkowski system $(ict, r, \theta, \phi)$, the
former can be written
\[
   ds^2=(icdt)^2+dr^2+r^2(d\theta^2
   +\text{sin}^2\theta\,d\phi^2),
\]
thereby retaining the Euclidian form.  So the metric
(\ref{E:line3}) becomes (since $R=1$)
\begin{equation}\label{E:line8}
   ds^2=(icdt)^2+a^2(t)[d\psi^2+\text{sin}^2\psi(d\theta^2
  +\text{sin}^2\theta\,d\phi^2)].
\end{equation}

The general theory of relativity is based on a
four-dimensional pseudo-Riemannian geometry.
With an extra spatial dimension, the natural
analogue of a 3-sphere is a 4-sphere or a
4-hyperboloid.  To obtain this analogue,
we would have to include (with $c=1$ again)
both $(idt)^2$ and $\text{sin}^2(it)$ or both
$(idt)^2$ and $\text{sinh}^2(it)$ in order to
retain the required form of the line element.

Since the extra spatial dimension is not
affected by the expansion, the scale factor
$a^2(t)$ remains in its original position.
However, the forms of the resulting metrics
require an extra term, here denoted by
$d\eta^2$:
\begin{equation}\label{E:line9}
   ds^2=d\eta^2+(idt)^2+\text{sin}^2(it)\{a^2(t)
   [d\psi^2+\text{sin}^2\psi(d\theta^2
  +\text{sin}^2\theta\,d\phi^2)]\}.
\end{equation}
and
\begin{equation}\label{E:line10}
 ds^2=d\eta^2+(idt)^2+\text{sinh}^2(it)\{a^2(t)
   [d\psi^2+\text{sin}^2\psi(d\theta^2
  +\text{sin}^2\theta\,d\phi^2)]\}.
\end{equation}
(Using the identities
$\text{sin}\,(it)=i\,\text{sinh}\,t$
and
$\text{sinh}(it)=i\,\text{sin}\,t$,
these line elements could also be written
with signatures \,\,$+----$\,\, or \, $-++++$.)
Eqs. (\ref{E:line9}) and (\ref{E:line10})
have the forms of a 4-sphere and a
4-hyperboloid, respectively.  The
respective spatial parts therefore
represent a four-dimensional boundary
of a five-dimensional ball and
hyperboloid.  Observe that $t$ has
become spacelike.

\section{Discussion}

The FLRW expanding closed Universe is
a 3-sphere.  Like the surface of an
expanding balloon, there is no edge
and every point has the appearance of
a center that all the other points on
the surface recede from.  So the best
way to describe a 3-sphere is a
three-dimensional boundary of a
four-dimensional ball.  There is no
edge and every point looks like the
center of the Universe.  The open
case leads to a 3-hyperboloid.

In this note we considered the
Euclidean form of the FLRW model
with $c=1$:
\begin{equation*}
   ds^2=(idt)^2+a^2(t)[d\psi^2+\text{sin}^2\psi(d\theta^2
  +\text{sin}^2\theta\,d\phi^2)]
\end{equation*}
or
\begin{equation*}
   ds^2=(idt)^2+a^2(t)[d\psi^2+\text{sinh}^2\psi(d\theta^2
  +\text{sin}^2\theta\,d\phi^2)].
\end{equation*}
Since the absence of a spatial edge is consistent
with observation, the extra spatial dimension
suggests an extension of the FLRW model that
results in the line elements (\ref{E:line9})
and (\ref{E:line10}).  These ideas lead to
the following possible interpretations:

\subsection{Interpretation 1}
The respective spatial parts in (\ref{E:line9})
and (\ref{E:line10}) represent a four-dimensional
boundary of a five-dimensional sphere or
hyperboloid.  Being a boundary, there is, once
again, no edge.  The difference is that the
boundary includes the $t$ in the \emph{original}
Lorentzian spacetime, which implies that the
Universe cannot have an edge, either spatial
or temporal.  The absence of a temporal edge
is consistent with Hawking's no-boundary
proposal.

The required form of the metric forced the
introduction of a new component in the
five-dimensional model,  Eqs. (\ref{E:line9})
and (\ref{E:line10}).  The original time $t$
is, as in Hawking's theory, an illusion,
making the five-dimensional time $\eta$ the
``real" time; moreover, the new time is
orthogonal to the original time.

\subsection{Interpretation 2}

Alternatively, one could view the
five-dimensional space as a mathematical
abstraction whose only purpose is to define
the four-dimensional boundary, just as a
four-dimensional fictitious sphere is used
to define the boundary of a 3-sphere.  With
this interpretation, there is no physical
second time component, leaving only ordinary
time, but there is still no temporal edge and
hence no beginning.
\\
\\

From the standpoint of the cosmological
principle, Interpretation 1 is probably
preferred.  Recall that according to this
principle, there is no such thing as a
special \emph{place}: everything looks
essentially the same in every direction
for every observer.  Yet strictly speaking,
everything should look the same at any
\emph{time}, as well.  Returning to $t$,
thanks to the Big Bang, this aspect of
the cosmological principle is lost: the
Universe would look different for any
two observers who are widely separated
in time, i.e., at respective times 
$t_1$ and $t_2$, where $t_1\ll t_2$.  
No such effect exists for the new 
time $\eta$, however, so that $\eta$ 
reestablishes the ``perfect 
cosmological principle."

\end{document}